# Heat Transport in Mesoscopic Systems

D.E. Angelescu, M.C. Cross and M.L. Roukes
Condensed Matter Physics
Caltech 114-36, Pasadena CA 91125

**Abstract**

Phonon heat transport in mesoscopic systems is investigated using methods analogous to the Landauer description of electrical conductance. A "universal heat conductance" expression that depends on the properties of the conducting pathway only through the mode cutoff frequencies is derived. Corrections due to reflections at the junction between the thermal body and the conducting bridge are found to be small except at very low temperatures where only the lowest few bridge modes are excited. Various non-equilibrium phonon distributions are studied: a narrow band distribution leads to clear steps in the cooling curve, analogous to the quantized resistance values in narrow wires, but a thermal distribution is too broad to show such features.

## 1 Introduction

The study of electronic transport in mesoscopic systems has uncovered many fascinating quantum aspects of resistance. The idea of relating charge transport in confined geometries to a quantum transmission problem, developed by Landauer in a series of papers starting forty years ago [1], has been an important unifying theoretical idea. The idea, in its modern understanding [2], can be expressed in terms of the formula for the conductance between two ideal electrodes

$$G = \frac{e^2}{h} Tr\ tt^+, \tag{1}$$

where $t$ is the transmission matrix between the electrodes and $e^2/h$ appears as the fundamental unit of conductance. In the case of an ideal long one-dimensional wire with no scattering Eq.(1) reduces to

$$G = N_c \frac{2e^2}{h}, \tag{2}$$

with $N_c$ the number of channels available for the transport, i.e. the number of transverse modes with energies below the Fermi energy of the electrodes. This leads to a quantized conductance and well defined steps in the measured resistance as the Fermi energy or the effective width of the wire is changed, for example using gate electrodes [3].

Implicit (and sometimes explicit) in the early work was the idea that boson excitations, such as phonons, could also be described by the same type of formalism. Recent experimental innovations in the construction of well characterized thermal reservoirs and thermal transport pathways makes a detailed analysis of the consequences of these ideas to heat transport in mesoscopic systems of considerable interest. In this paper we present results in this direction.

## 2 Experimental Systems

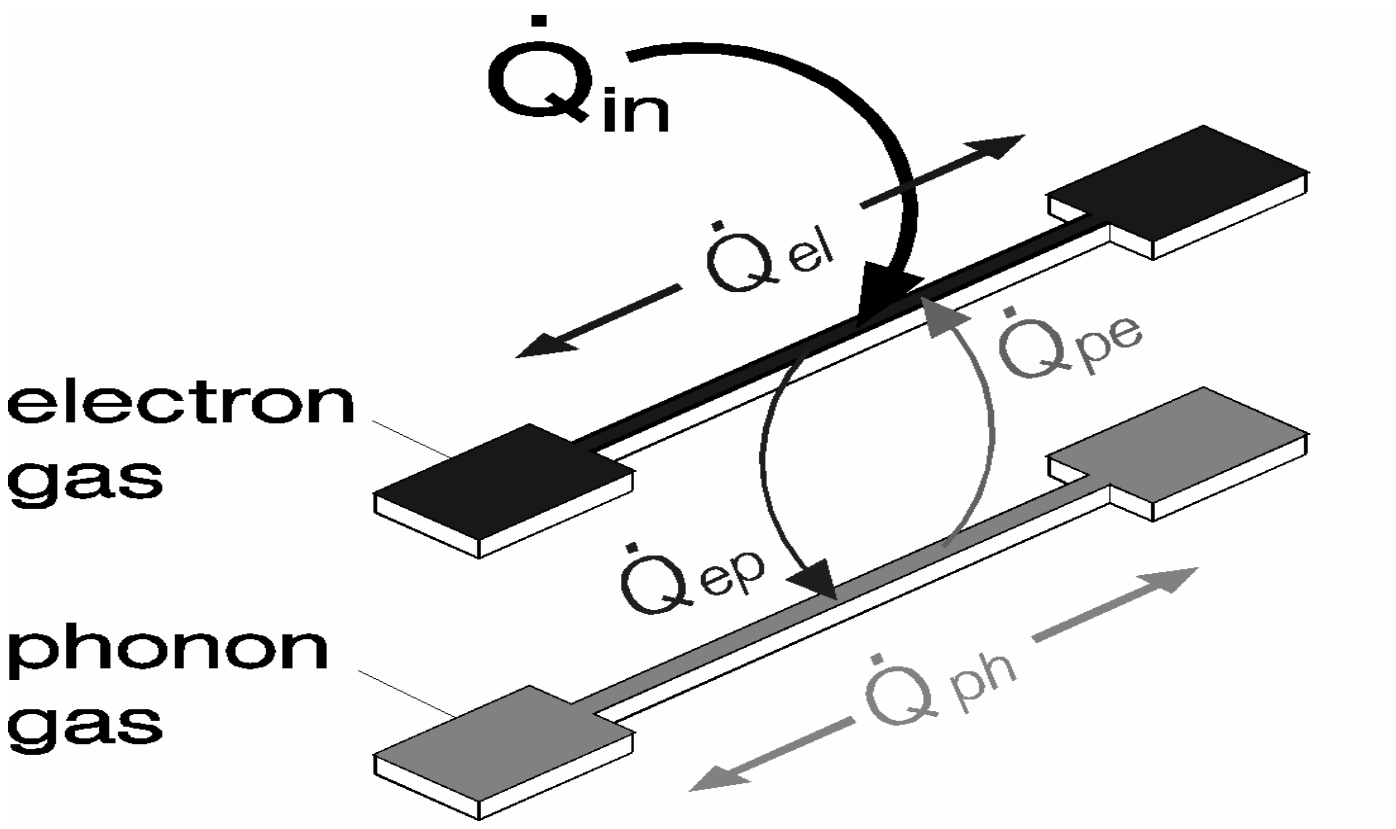

Figure 1: Conceptual view of an electron heating experiment. Joule heat $\dot{Q}_{in}$ is introduced to the electrons. Some of this is transferred to the phonons by the e-p interaction $\dot{Q}_{ep}$; this heats the phonons, which then cool by phonon diffusion $\dot{Q}_{ph}$ and by returning some of this heat to the electrons $\dot{Q}_{pe}$. However heat also flows to the contacts via electron diffusion $\dot{Q}_{el}$.

To date, relatively few groups have attempted experimental explorations of phonon transport in mesoscopic structures. For most that have, the measurement strategy has been to induce electron heating in wires that are thermally-decoupled from the environment, either through physical suspension or through intentionally poor thermal contact (e.g. Kapitsa boundary resistance). What is actually measured is the electron energy loss rate, which at low temperatures and in the absence of phonon heating is determined by the phonon emission rate from hot electrons [4]. The underlying premise in this approach is that if phonon transport is sufficiently impaired hot phonons will result—and the *electronic* properties such as resistance will give information on the *phonon* heat pathways. However, the actual picture is somewhat more complicated (figure 1). The thermally-decoupled, conducting samples employed in these studies comprise both electronic and phononic thermal conductors, and the electron cooling path by electron diffusion down the wire to the cooler contacts "short-circuits" the phononic pathway of interest. To deduce the phononic contribution in this situation requires rather involved modeling of heat flow in these structures, with tacit assumptions about the scattering rates occurring within mesoscopic systems in the non-equilibrium steady state.

The earliest attempts to study phonon transport in mesoscopic systems began with suspended metal wires in 1985 [5], and culminated in investigations of arrays of suspended, monocrystalline n+ *GaAs* wires [6]. Electron diffusion overwhelmed the indirect energy pathway involving phonons in these samples, hence little evidence of a phononic contribution to cooling could be observed. In the later work at higher currents, analysis using a coupled electron/phonon model provided some evidence for the onset of a small contribution to electron cooling via a phononic pathway, but it also indicated that the phonons were not appreciably heated in the process.

Electron heating experiments in both suspended and supported wires of a polycrystalline *AuPd* alloy first initiated in 1985, led to an intriguing report of the observation of acoustic waveguide modes in 1992 [7]. The samples manifesting these unusual features were anchored to the substrate. Nonetheless, it was argued that phonon confinement could occur due to acoustic mismatch at the substrate/metal film interface. In the intervening time the search by others for similar phenomena has proven elusive [8]. The results in the *AuPd* system were observed in a temperature regime where the thermal smearing of the electron distribution function ($4k_BT \sim 2meV$) was comparable or larger than the sharpness of the features seen in the data ($\sim 0.3meV$). A mechanism remains to be elucidated which would enable electron-phonon processes involving a smeared Fermi surface to manifest features with the sharpness of those observed. It would appear necessary to postulate the generation of a non-equilibrium phonon distribution function with spectral features far narrower than those occurring in thermal equilibrium (see section 4.2.2). These topics remain to be explored experimentally.

The recent work of Tighe, Worlock and Roukes [9] in *GaAs* heterostructures was motivated by a desire to configure monocrystalline nanostructures for *direct* phonon thermal conductance measurements. Conceptually, the simplest approach for such measurements would be to thermally clamp one end of an insulating nanostructured beam under study while providing heat input $\dot{Q}$ to the other thermally-isolated end. The thermal conductance for small heat input is then $K = \dot{Q}/\Delta T$, where $\Delta T$ is the temperature difference between the clamped and thermally isolated ends. Fabrication of *nm*-scale structures in such a configuration is awkward: it is difficult to support a thermally-isolated end without placing undue strain on the suspended beam. Instead, the approach depicted in figure 2 is taken in which a phonon "cavity", formed from insulating *GaAs* in the shape of a thin plate, is suspended above the substrate by insulating *GaAs* beams. To measure the thermal conductance of these nanostructured beams, the cavity is Joule heated by a *source* transducer patterned above it. The cavity then cools through the long, narrow, monocrystalline insulating *GaAs* bridges that suspend it. The rise in cavity temperature is measured using a second *sensor* transducer. This configuration allows direct measurement of the parallel thermal conductance of the four *nm*-scale support beams. Through this approach it has now become possible to pattern separately the *mesoscopic insulators*, which determine the thermal transport, and the *transducers* which are used to induce and measure a response. Details of the measurement technique and the results obtained may be found in reference [9]; in the present context these devices serve to illustrate that an experimental path into the regime of mesoscopic

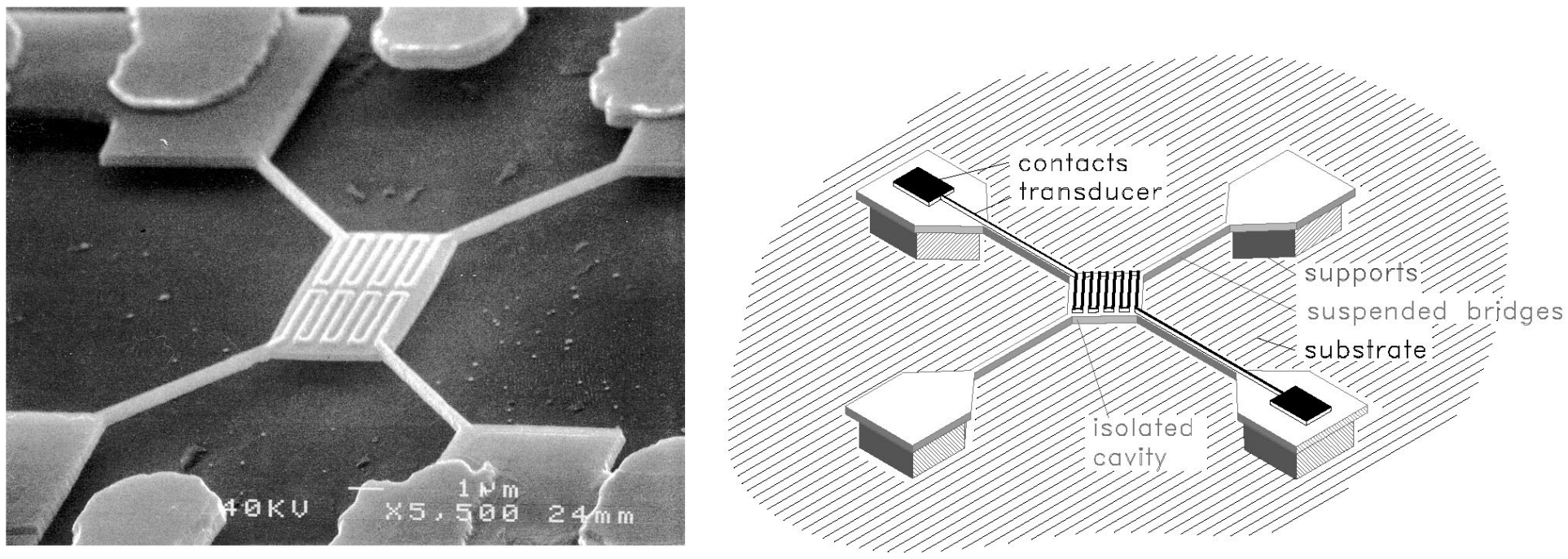

Figure 2: Suspended, monocrystalline device enabling direct thermal conductance measurements on nanostructures. (Left). Electron micrograph showing suspended, semi-insulating *(i) GaAs* plate ($3 \times 3$ $\mu m$ $\times$ $300 nm$ thickness), forming a quasi-isolated phonon cavity, that is suspended $1\mu m$ above the substrate by four $5.5\mu m$ long *i-GaAs* bridges (cross section $200nm \times 300nm$). An integral pair of $n+$ *GaAs* resistive transducers ($120nm$ linewidth, $150nm$ thickness), meander above, but are in epitaxial registry with, the underlying *GaAs* cavity. (Right). Schematic diagram showing principal components of the suspended device: isolated cavity, suspended bridges, and the supports anchoring the sample to the substrate, which together comprise the external reservoir. (After Tighe et al.[9])

thermal transport now exists.

# 3 The Ideal One-dimensional Heat Conductor

Motivated by the experimental geometry of figure 2 we consider the situation pictured in figure 3. A small thermal mass, the *cavity*, at temperature $T_1$ is thermally isolated from the environment except for contact through a narrow *bridge* to the *reservoir* at temperature $T_2$. We will study the heat transport for the case where $T_2 = 0$ modeling cooling experiments where the reservoir is at a much lower temperature than the cavity, and also the case with the temperature difference $T_1 - T_2$ small compared to the mean temperature corresponding to measurements of thermal conductance. The simplest model of the thermal transport is given by assuming the modes in the bridge are populated with an equilibrium thermal distribution given by the appropriate temperature—the temperature of the cavity for the right moving phonons, and that of the reservoir for the left moving phonons. This corresponds to an "adiabatic" coupling of the modes in the bridge and reservoirs, as would be expected for a smooth, tapered junction. (The experimental geometry with the bridges connected to the corners of the square cavity might be expected to approximate this situation quite well.) In this case we are calculating the heat transport as a property of the bridge alone, and are neglecting scattering (due to mode-mismatch) at the cavity-bridge interface.

The calculation of the heat transport proceeds as in the calculation of the ideal electrical conductance Eq.(2) except that we are interested in energy transport, rather than number transport, and of course the thermal distribution is given by the Bose distribution $n(\omega)$ rather than the Fermi distribution. If we first look at the transport by the right moving phonons, the energy flux is

$$H^{(+)} = \frac{1}{2\pi}\sum_m \int_0^\infty dk\, \hbar\omega_m(k)\, n(\omega_m(k))\, v_{gm}(k)\,, \tag{3}$$

where $k$ is the wave vector along the bridge, $\omega_m(k)$ is the dispersion relation of the $m$th discrete mode of the bridge, and $v_{gm} = d\omega_m(k)/dk$ is the group velocity. Transforming the integral to an integral over frequencies yields an expression for the heat transport by right moving phonons that can be written as a sum of mode contributions $H^{(+)} = \sum_m H_m^{(+)}$ with

$$H_m^{(+)} = \frac{1}{2\pi}\int_{\omega_m}^\infty d\omega\, \hbar\omega_m(k)\; n(\omega_m(k))\;, \tag{4}$$

where $\omega_m$ is the *cutoff frequency* of the $m$th mode, i.e. the lowest frequency at which this mode propagates. (We have assumed the $m$th mode propagates to arbitrarily large frequencies. If a particular mode only propagates over a finite band of frequencies, the upper limit of the integral will be replaced by $\omega_m^{\max}$.) Using the Bose distribution for $n(\omega)$ leads to the expression

$$H_m^{(+)} = \frac{(k_B T_1)^2}{h}\int_{x_m}^\infty \frac{x}{e^x - 1}dx, \tag{5}$$

where $x_m = \hbar\omega_m/k_B T_1$ and again a maximum frequency to the phonon dispersion would lead to $\hbar\omega_m^{\max}/k_B T_1$ appearing as the upper limit.

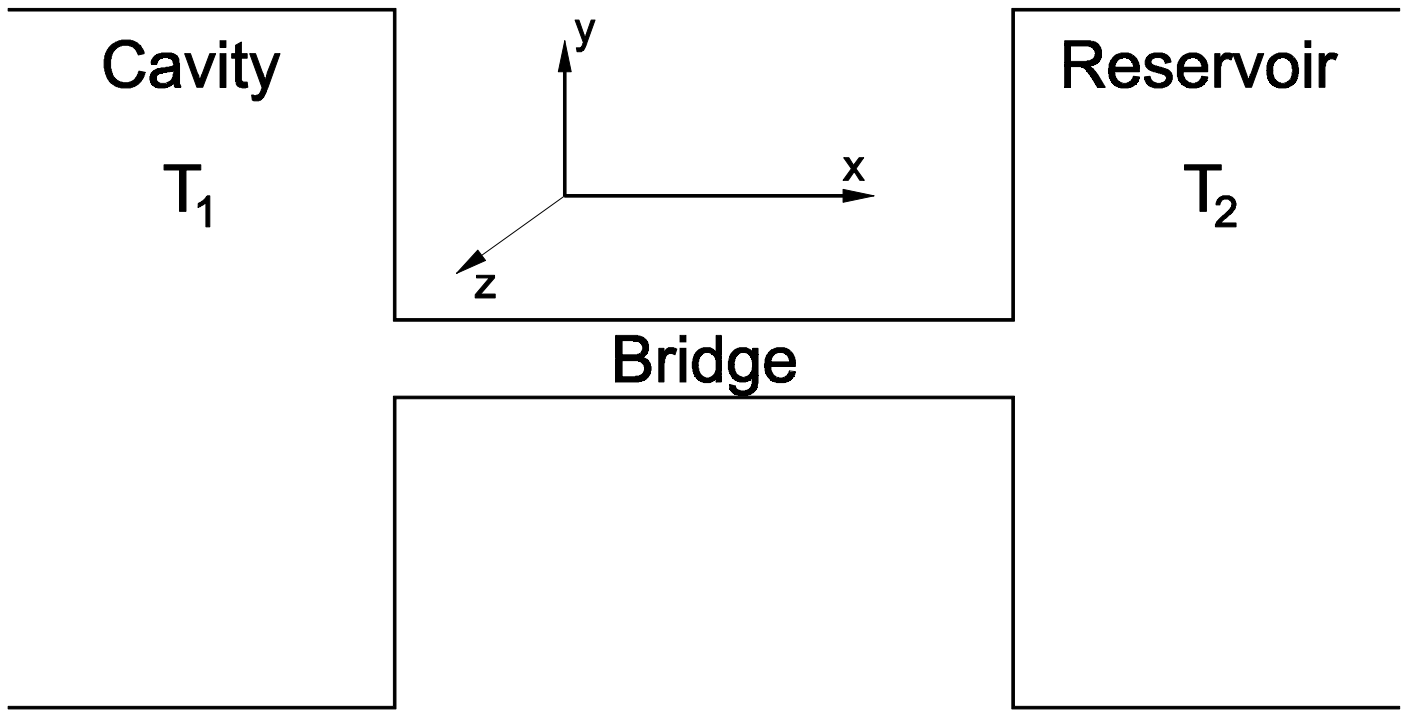

Figure 3: Schematic diagram of geometry for heat transport experiment

The heat transport is given by summing $H_m^{(+)}$ over the modes $m$, and if the reservoir is at a non-zero temperature $T_2$, subtracting the analogous expression for $\sum_m H_m^{(-)}$ for the left moving phonons given by Eq.(5) (but with $T_2$ replacing $T_1$). In a thermal conductance measurement the cavity and reservoir are maintained at a temperature difference small compared to their temperatures, and the thermal conductance is given by the temperature derivative of $H^{(+)}$

$$K = k_B \sum_m \frac{k_B T}{h} \int_{x_m}^{\infty} \frac{x^2 e^x}{(e^x - 1)^2} dx. \tag{6}$$

Equation (6) plays the role of a "universal" phonon conductance in direct analogy with the expression for the electronic case (2). Note that this result is *independent of all details of the dispersion curve* except the cutoff frequencies $\omega_m$. (The same remark applies to the result for $H^{(+)}$). This simple result arises, as in the electronic case, because the density of states factor in the frequency integral is cancelled by the group velocity. Equations (5) and (6) will play a central role in our discussion of heat transport in mesoscopic systems.

# 4 Scalar Model

## 4.1 Adiabatic Coupling

To investigate the experimental consequence of Eqs.(5) and (6) we consider a *scalar* model of elasticity where the modes are given by a scalar field $\phi$ which satisfies the wave equation

$$\frac{\partial^2 \phi}{\partial t^2} - c_i^2 \nabla^2 \phi = 0. \tag{7}$$

We allow for the polarizations of the elastic wave (2 transverse, 1 longitudinal) simply by summing over three independent modes with the wave speeds $c_i$ given by $c_t$ (transverse) or $c_l$ (longitudinal): in this scalar theory we ignore mode mixing effects that would occur at boundaries and interfaces. Boundary conditions on $\phi$ of zero normal derivative are assumed, corresponding to free surfaces. (This allows the propagation of acoustic modes $\omega \propto k$ along the bridge, which is a partial modeling of the situation for elastic waves.) For the temperatures of interest the wavelength of the excited modes is large compared to the atomic scale, so that the linear dispersion relation in Eq.(7) is adequate. These simple assumptions lead to waveguide modes

$$\omega_{mn}(k) = \sqrt{\omega_{mn}^2 + c_i^2 k^2}, \tag{8}$$

where the cutoff frequencies are given by

$$\omega_{mn} = c_i \pi \sqrt{\left(\frac{m}{w}\right)^2 + \left(\frac{n}{d}\right)^2} \tag{9}$$

with $w$ the width and $d$ the depth of the bridge. For simplicity we will present results for a single sound speed, all $c_i = c$.

With the mode structure defined, we can now calculate the cooling rate of a cavity in contact with a zero temperature reservoir, and the thermal conductance. For ease of presentation we show results for the thin limit $\lambda_{th} \equiv hc/k_B T \gg d$ in both the bridge and reservoirs, so that there are no modes excited across the depth of the material i.e. effectively a two dimensional situation. Here $\lambda_{th}$ is the thermal wavelength demarcating the transition between thermally populated and unpopulated modes. We will use the notation $\omega_m$ for $\omega_{m0}$. To calculate the cooling rate of the cavity we then use the 2-dimensional expression for the Debye thermal capacity:

$$C = S \frac{9\zeta\ (3)\ k_B^3}{\pi\ (\hbar c)^2} T^2 \tag{10}$$

with $S$ the area of the cavity.

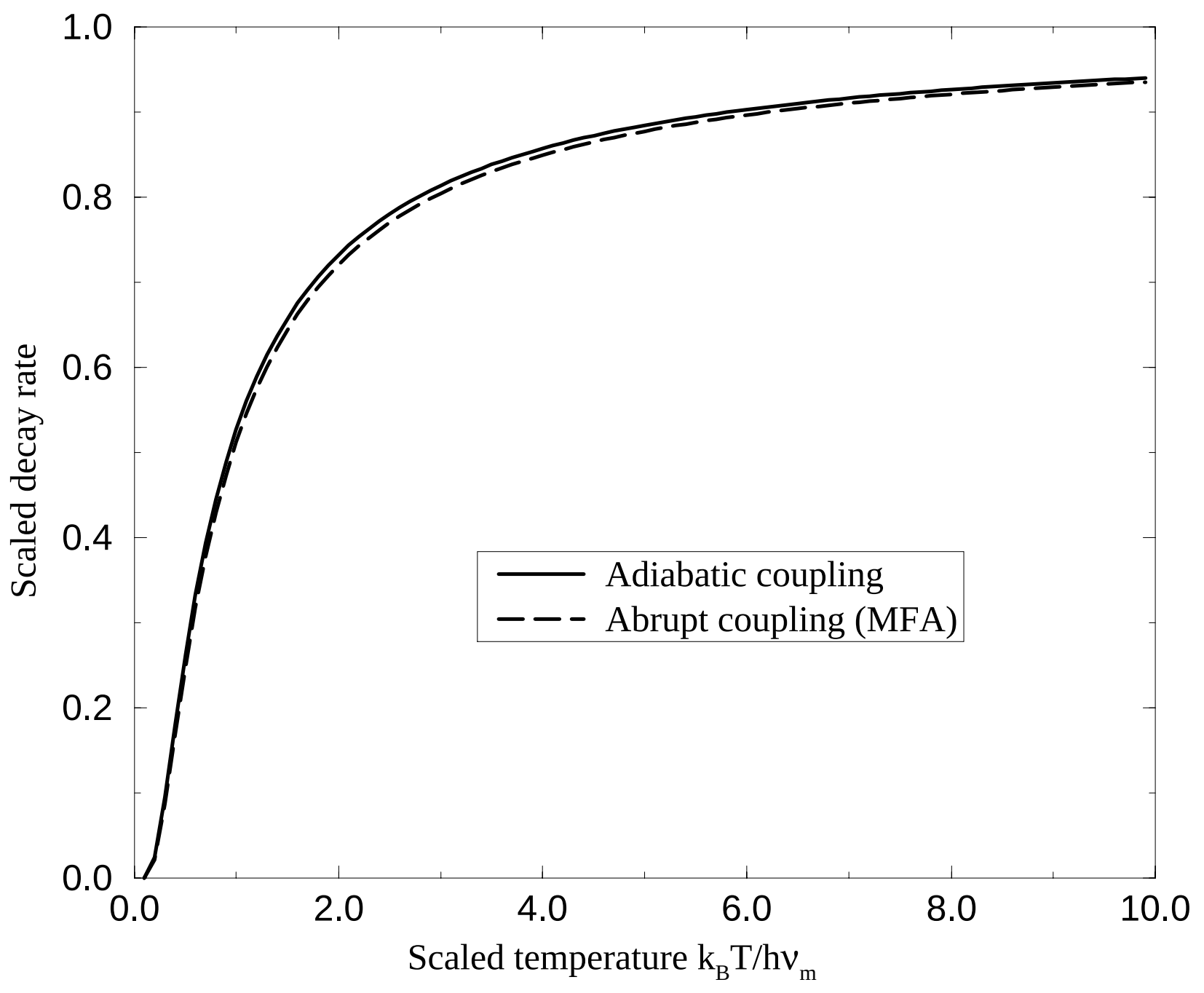

Figure 4: Cooling rate due to single bridge mode as a function of reduced temperature. The vertical axis is scaled to unity at large temperatures. The solid curve is the result for adiabatic coupling, the dashed curve for abrupt coupling in the "mean field approximation" (see appendix).

The cooling rate $|dT/dt|$ of a cavity at temperature $T$ connected via the bridge to a reservoir at zero temperature given by a single bridge mode $m$ is shown in figure 4. At high temperatures the energy in each bridge state is given by classical equipartition, and the heat transport rate is proportional to $T^2$, the same temperature dependence as the two dimensional specific heat. For temperatures below $k_B T = \hbar\omega_m$ the heat transport falls with the mode occupation, eventually falling exponentially at low temperatures. However the cutoff is quite smooth, unlike the case of electrical transport where the Fermi function can lead to sharp steps. (If instead we had considered a three dimensional cavity, with a $T^3$ thermal capacity, the cooling rate per mode would show a broad peak at around $k_B T = \hbar\omega_m$.)

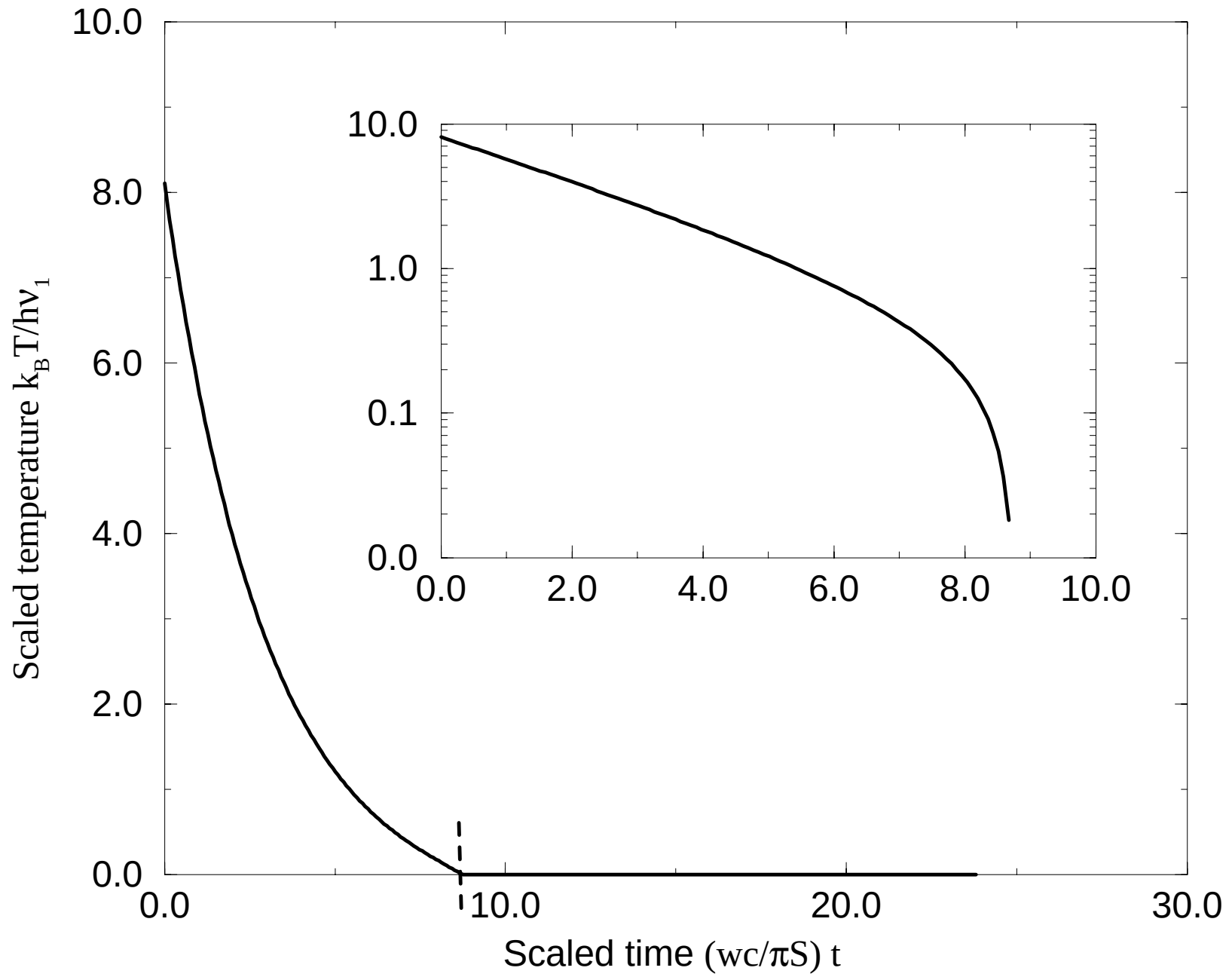

Figure 5: Temperature decay rate for cavity connected to a reservoir at zero temperature in the approximation of adiabatic coupling to the bridge. The inset shows the same curve on a log-linear plot.

Summing over the waveguide modes given by Eq.(9) yields the prediction for the cooling curve $T(t)$, figure 5. Note that the cooling curve is quite smooth, not showing any features of the discrete mode structure of the transport pathway due to the broad nature of the cutoff given by the Bose distribution (figure 4). This can be compared to the Fermi case, where the ability to tune two different parameters separately—the Fermi energy through the gate voltage, and the width of the thermal broadening through the temperature—allows the sharp step structure to be measured. The important point is that in the phonon case the temperature sets both the range of transport modes that are effective, and also the thermal broadening of the mode cutoff, and in this case the width of the distribution function is sufficient to smear out the quantized signature of the transport.

An interesting feature of the cooling curve is that at very low temperatures, where only the acoustic $m = 0$ mode is excited, the equation for the cooling becomes

$$\frac{dT}{dt} = -\frac{\hbar c^2 \zeta\,(2)}{6 k_B \zeta\,(3)} \frac{1}{S}, \tag{11}$$

i.e. the temperature decreases linearly in time, and goes to absolute zero at a finite time of order $\pi S/wc$! (In the real case, with a reservoir at a nonzero temperature, the cavity temperature would equilibrate at the reservoir temperature.) This result no longer holds for non-adiabatic coupling, since the very long wavelength modes excited at low temperatures are particularly sensitive to the precise details of the intermode coupling, and their contribution to the heat transport is reduced.

The prediction for the thermal conductance as a function of temperature is shown in figure 6. Again there is no signature of the discrete mode structure contributing to this curve. At high temperatures the number of modes contributing to the thermal conductance Eq.(6) grows proportionately with the temperature, so that $K \propto T^2$. Since the cavity heat capacity also varies as $T^2$ the exponential relaxation rate of small temperature differences between cavity and reservoir $K/C$ becomes independent of temperature at high temperatures. This is shown in the inset to figure 6.

## 4.2 Non-equilibrium phonon distributions

### 4.2.1 Relaxation with no internal equilibration.

In the calculations in the previous section, it is assumed that the phonons in the cavity have sufficient time to rethermalize during the cooling process. However, since anharmonic effects should be small at low temperatures, this may not be the case, and it is interesting to consider the opposite limit where there is no rethermalization. In this case we can consider each frequency band $\omega \to \omega + \delta\omega$ separately. The spectral energy density in the cavity is $E\,(\omega) \sim n\,(\omega)\,\omega^2$ where $n$ is now the non-equilibrium distribution and the $\omega^2$ is from the 2-dimensional density of states. The heat transport per unit frequency is $H^{(+)}\,(\omega) \sim n\,(\omega)\,\omega N_c\,(\omega)$ where $N_c$ is the number of bridge modes accessible at frequency $\omega$ (i.e. with cutoff frequency less than $\omega$), since each accessible mode contributes the same amount to the heat current. Thus each frequency band will relax exponentially with a time constant $\tau\,(\omega)$. For the narrow range of frequencies between adjacent bridge cutoff frequencies, where $N_c$ is constant, the decay rate for the lower frequencies will be larger than for the upper frequencies ($\tau\,(\omega) \propto \omega$ for fixed $N_c$). But over wider ranges of frequencies (large compared to the bridge mode spacing), since $N_c$ grows on average linearly with increasing frequency, the time constant $\bar{\tau}\,(\omega)$ will be roughly frequency independent. Thus the total energy is expected to decay as a single exponential, and this is confirmed by direct simulations. In addition the envelope of $n(\omega)$ will decay exponentially at this rate, even though considerable fine scale structure on a scale of the bridge mode cutoff frequency separation develops (figure 7).

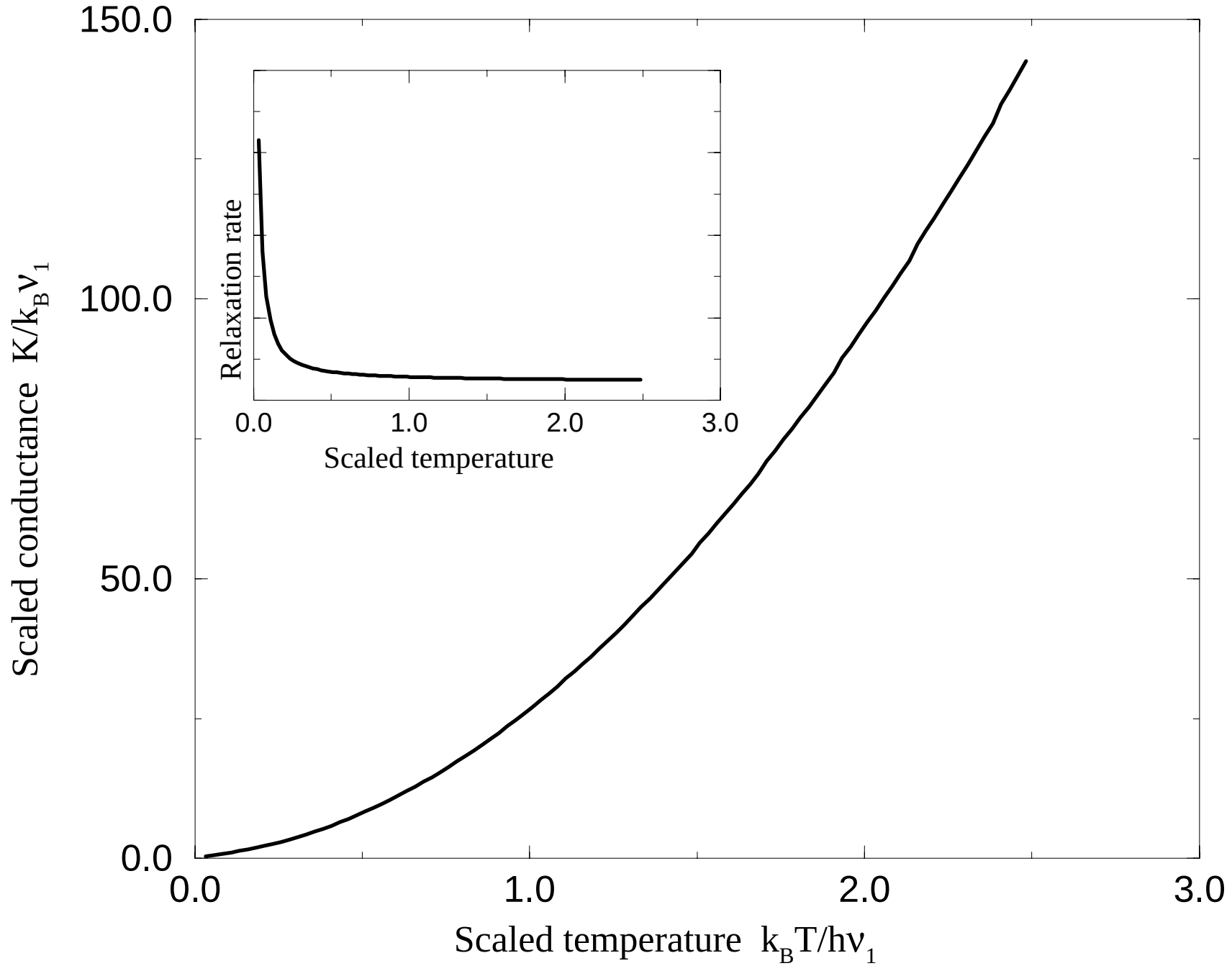

Figure 6: Thermal conductance $K$ (scaled by $k_B \nu_1$) as a function of temperature (scaled to $h\nu_1/k_B$), with $\nu_1$ the first bridge mode cutoff frequency. The inset shows the resulting temperature relaxation rate $K/C$ with $C$ the cavity thermal capacity (arbitrary units).

We could also imagine situations where reduced transmission rates of particular modes change this result. For example a beam with modulated width could be manufactured to induce a band gap in the lowest bridge mode so that the transport by this mode is significantly reduced by an effective transmission coefficient $\mathcal{T}_0$. (There would also be gaps in the other modes at the wave number of the modulation, but these are expected to be smaller than in the lowest mode which should be perturbed more strongly by the geometry.) In this case all frequency bands other than $\omega < \hbar c\pi/w$ should relax with the time scale $\bar{\tau}$, whereas this lowest band should relax with a longer time constant $\bar{\tau}/\mathcal{T}_0$. This slower exponential should become apparent at long times, as shown in figure 8.

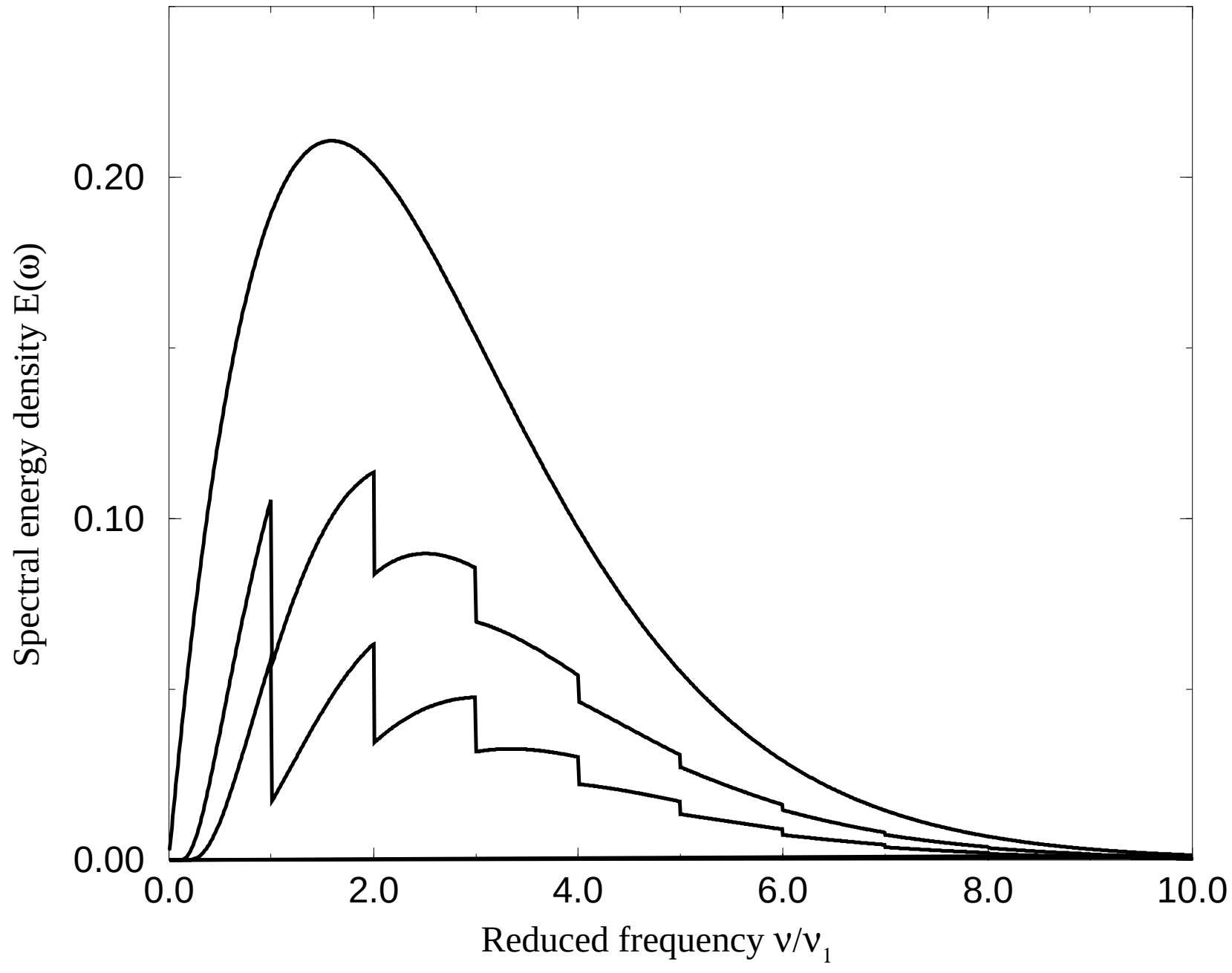

Figure 7: Spectral energy density $E(\omega)$ for three different times in the relaxation from a thermal distribution with no rethermalization in the cavity. The initial temperature corresponds to $h\nu_1/k_B$ with $\nu_1$ the lowest bridge cutoff frequency. The subsequent plots are at times 0.014 and 0.028 measured in units of $\pi S/wc$.

#### 4.2.2 Narrow band distribution

Since a thermal distribution appears to be too broad in most situations to give any signature of the mode quantization in heat transport, it is interesting to look at other, nonthermal distributions. We could imagine, for example, setting up a nonthermal phonon distribution in the cavity using superconducting junctions, or even using a modulated bridge with a band gap in the phonon spectrum to selectively filter the cavity phonons. In addition, our results for a thermal distribution give no signatures of the discrete phonon modes such as seen in the electron cooling experiments [7], and, although the geometry there is quite different from the one we consider, it is of interest to see whether a nonthermal phonon distribution might recover the signature of discrete modes.

We again model the coupling as adiabatic, so that the bridge modes are populated with the same distribution function as the cavity. We assume a distribution function of the form $n(\omega) = A[\theta(\omega - (\omega_0 - \Delta\omega/2)) - \theta(\omega - (\omega_0 + \Delta\omega/2))]$, *i.e.* a square band centered at $\omega_0$ and of width $\Delta\omega$. It is obvious that a mode with dispersion relation $\omega_m(k)$ and cutoff frequency $\omega_m = \omega_m(0)$ does not contribute to the heat transport if $\omega_m > \omega_0 + \Delta\omega/2$, since in that case the mode is not populated. Using the expression for $n(\omega)$ in Eq.(4) leads to the heat transport by a single mode:

$$H_m^{(+)} = \frac{A\hbar}{2\pi} \int_{Max(\omega_m,\omega_0-\Delta\omega)/2}^{\omega_0+\Delta\omega/2} \omega d\omega. \tag{12}$$

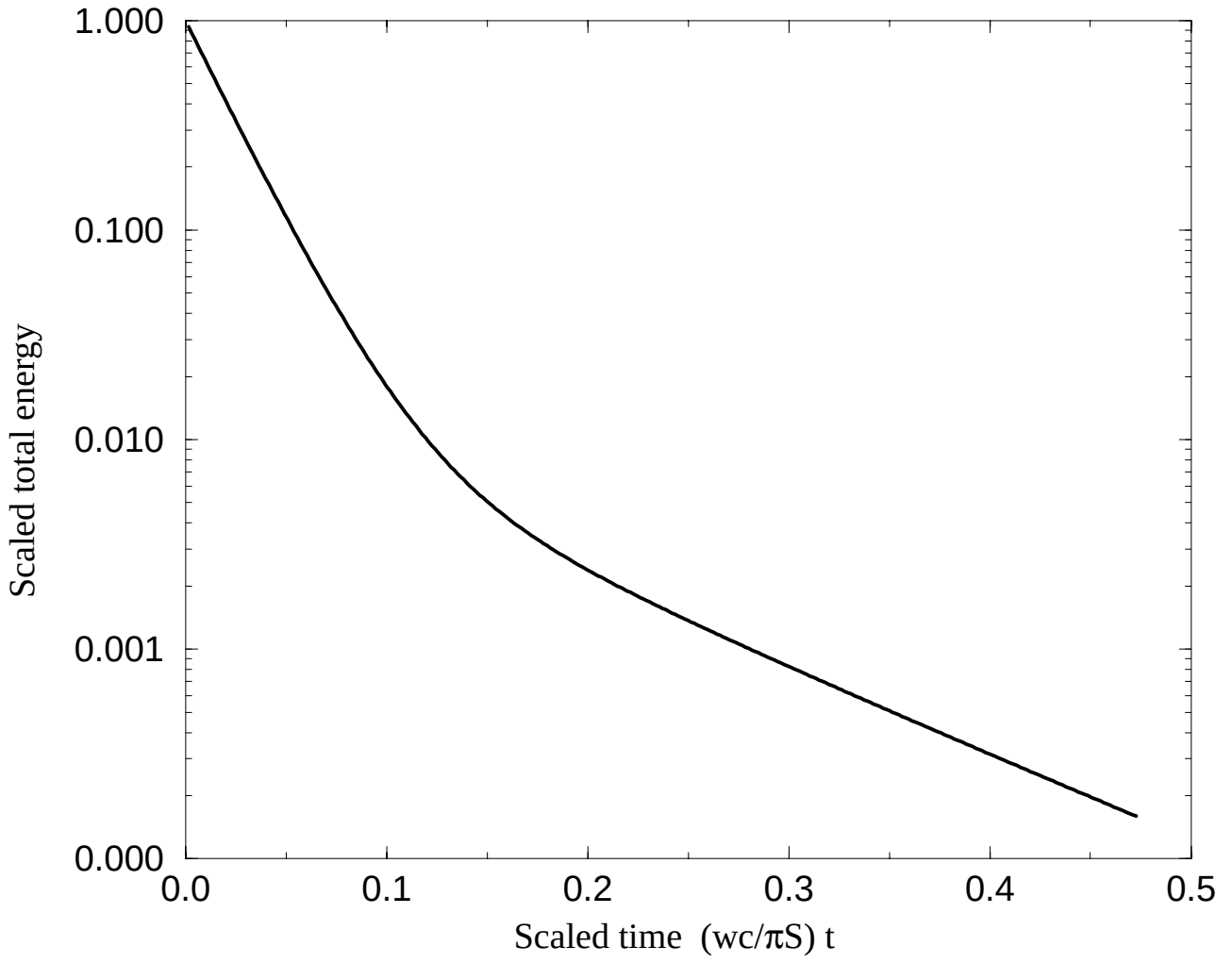

Figure 8: Decay of the total energy when the transmission into the lowest bridge mode is reduced to 0.2. The initial temperature corresponds to $3h\nu_1/k_B$ with $\nu_1$ the lowest bridge mode cutoff.

This formula predicts a constant heat transport rate equal to $\frac{A\hbar}{2\pi}\omega_0\Delta\omega$ for $\omega_0 > \omega_m + \Delta\omega/2$, and 0 for $\omega_0 < \omega_m - \Delta\omega/2$. Thus, when the central frequency of the narrow band distribution passes through a mode cutoff frequency, the transport rate shows a "step" increase with a step sharpness, defined as the ratio between the size and the width of the step, given by $\frac{A\hbar}{2\pi}\omega_0$. Note that the apparent sharpness increases with the frequency. The decay rate of the total energy using the scalar model of phonons in figure 9 shows the expected result that a narrow band phonon distribution uncovers the effects of the discreteness of the bridge modes.

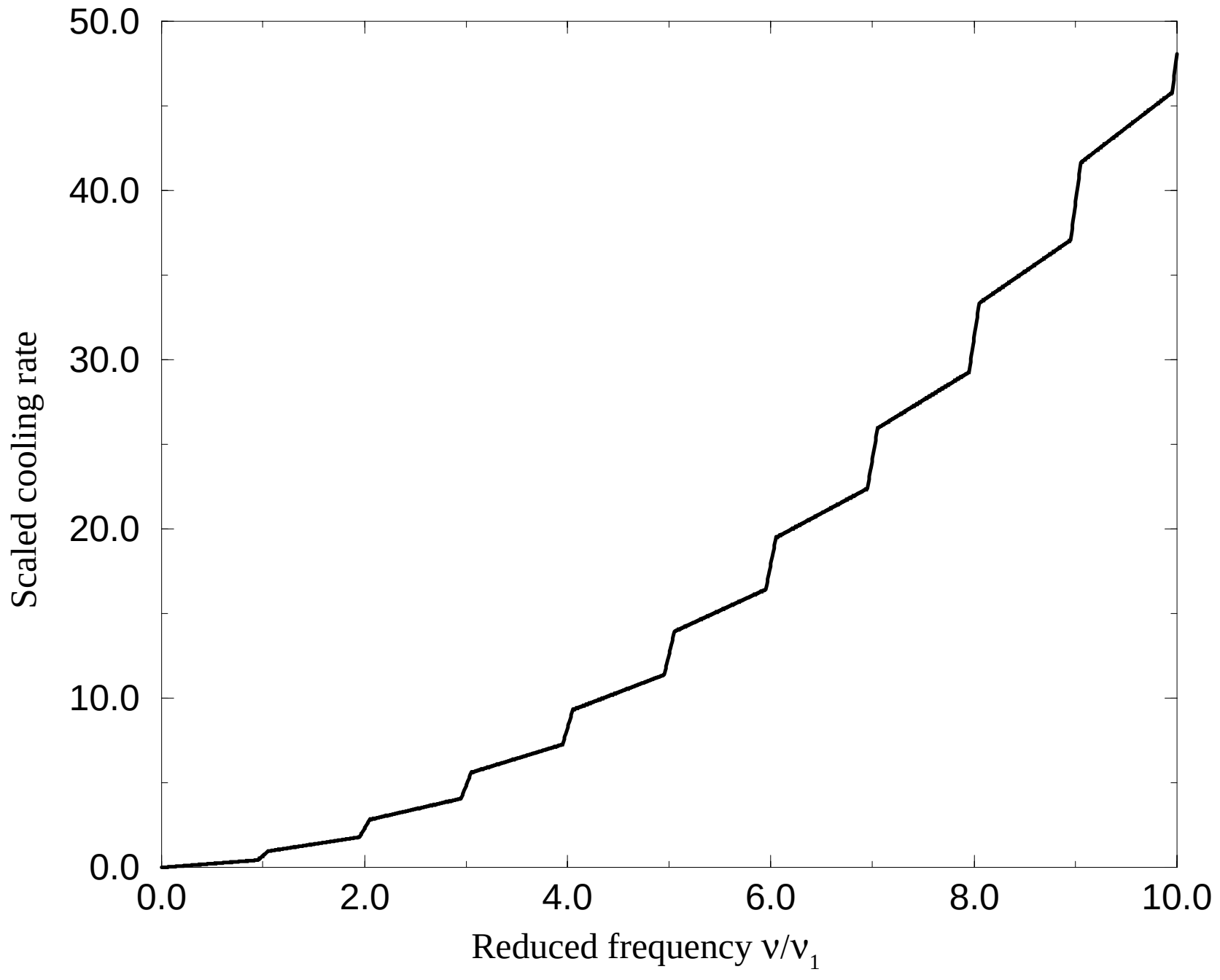

Figure 9: Cooling rate (rate of energy loss) as a function of the central frequency $\nu$ of a narrow band distribution, scaled to the first bridge mode cutoff frequency $\nu_1$. The width of the distribution $\Delta\nu$ was $0.1\nu_1$.

### 4.3 Non-adiabatic coupling

In general due to imperfect coupling of the cavity modes to the bridge modes, the population of the bridge modes will not be the same as the cavity distribution under conditions of thermal transport, i.e. when $T_1 \neq T_2$. The full expression for the thermal transport by the mode $m$, in the spirit of the Landauer formula, will involve a transmission probability $\mathcal{T}_m(\omega)$ which takes into account the imperfect transmission at the junction between the bridge and the reservoirs. We can estimate the importance of the geometry of the cavity-bridge junction by considering the worst case, that of an abrupt junction (as sketched in figure 3). In the experiment, this would correspond to the bridges meeting the cavity along a side, rather than at a corner. We show in the appendix that the mode thermal transport in this case can be written

$$H_m^{(+)} = \frac{1}{2\pi} \int_{\omega_m}^{\infty} \frac{\hbar\omega}{e^{\beta\hbar\omega} - 1} \mathcal{T}_m(\omega)\, d\omega, \qquad (13)$$

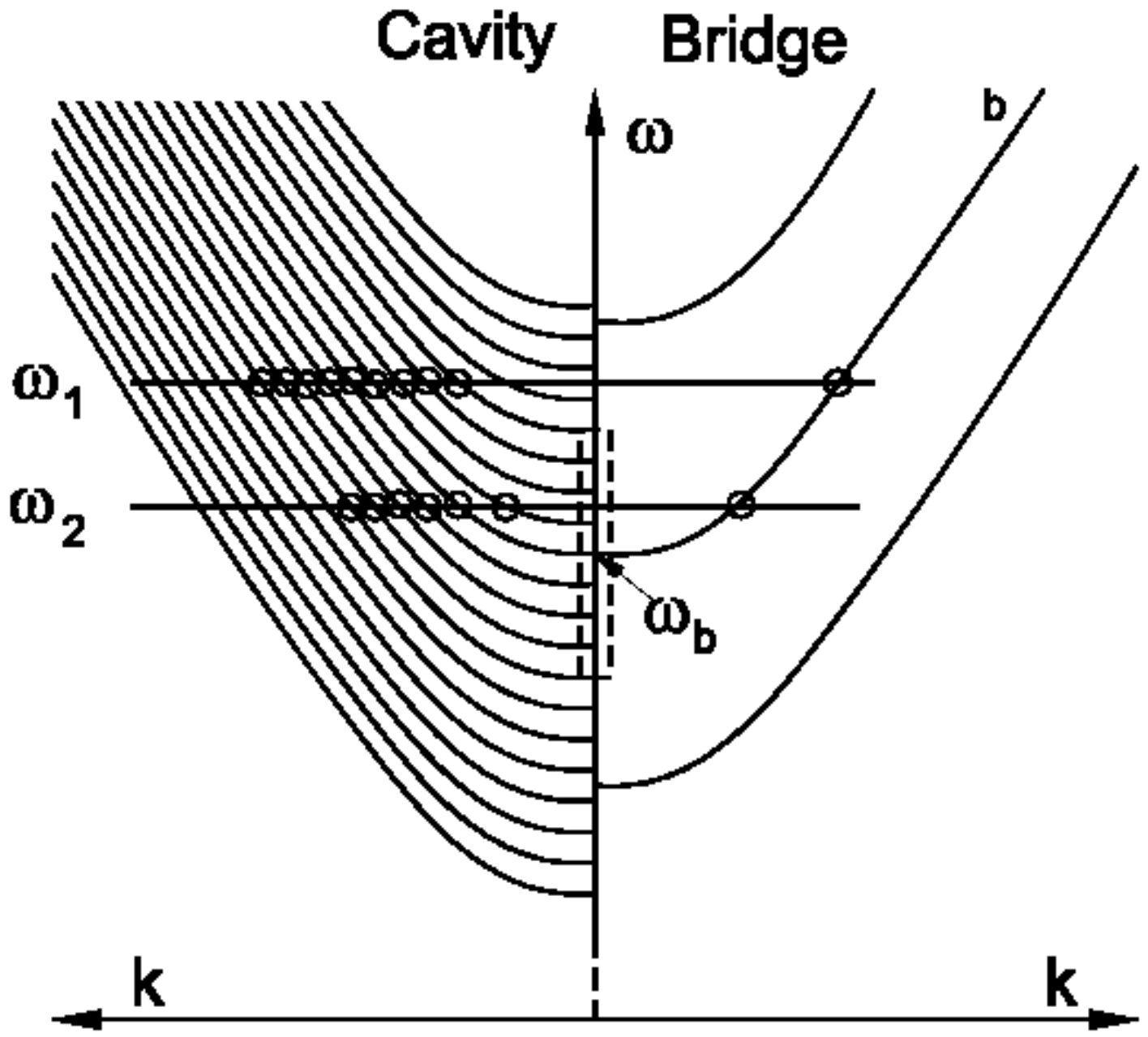

Figure 10: Cavity and Bridge Modes. The modes at two frequencies $\omega_1$ and $\omega_2$ that couple to a particular bridge mode $b$ are marked with a circle. The dashed square outlines the cutoff frequencies of the modes that strongly couple. The coupling is only between modes of like parity, and only modes of one parity are shown in the figure.

where $\mathcal{T}_m(\omega)$ can be calculated following the work of Szafer and Stone [10] on the analogous electronic problem. Indeed a simple analytic approximation scheme they suggest produces values for $\mathcal{T}_m(\omega)$ that are in good agreement with the numerically calculated results, except for low temperatures where only the lowest few bridge modes are active.

The scheme for calculating $H^{(+)}$ is to populate all the modes in the cavity at each energy $\hbar\omega$ (which are the modes that can couple to the $m$th bridge mode at the same energy) with the cavity distribution $n(\omega)$ (e.g. thermal), and then to calculate the total energy flux in the bridge from the transmitted part of the full wave functions. The key idea explained by Szafer and Stone, is that because of the properties of the overlap integral of the transverse spatial dependences of the cavity and bridge modes, a given (say, even) bridge mode will couple strongly only to (even) cavity modes with cutoff frequencies within $\pm\frac{1}{2}\Delta\omega$ of the bridge mode cutoff frequency where $\Delta\omega = 2\pi c/w$ is the bridge (even) mode cutoff spacing. Similar results hold for the odd modes. Then a sum rule, and the fact that the *average* transmission probability of the modes that do couple (see figure 10) rapidly approaches unity as $\omega$ increases above the bridge mode cutoff frequency, gives a $\mathcal{T}_m(\omega)$ that is close to a step function (see figure 11). This was pointed out by Szafer and Stone, and the argument is explained in more detail in the appendix. The approximation to a step function is very good for large bridge mode indices $m$, but becomes poorer as $m \to 0$. Thus, after performing the frequency integral, we find that the single mode thermal transport is, to quite high accuracy, unchanged from the adiabatic coupling limit at high temperatures ($\lambda_{th} = hc/k_B T \ll w$). An important result however is that for the $m = 0$ mode, $\mathcal{T}_0(\omega) \propto \omega^3$ for small $\omega$ (see appendix). This means that at low temperatures the cooling rate is determined by $dT/dt \propto T^3$ rather than a constant as in the adiabatic calculation, so that $T \sim t^{-1/2}$ in the low temperature limit. Similarly at low temperatures we have for the thermal conductance $K \propto T^4$ rather than the linear low temperature dependence predicted by Eq.(6). It should be noted however that these results depend on the scalar model of the elastic waves and on the detailed nature of the dispersion curves at low frequencies.

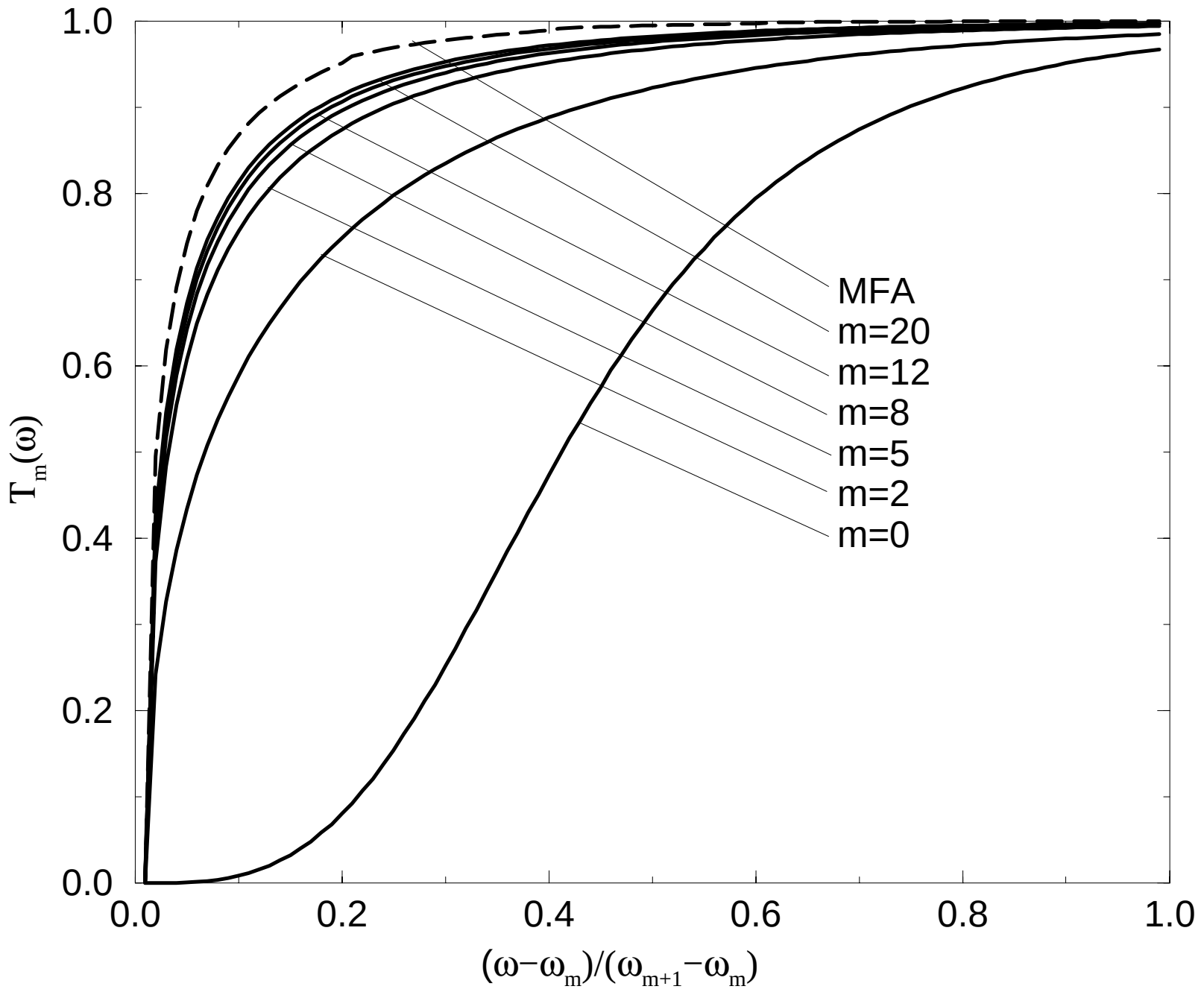

Figure 11: Transmission probability $\mathcal{T}_m$ for different bridge modes $m$ as a function of frequency $\omega$ above the mode cutoff frequency $\omega_m$. The frequency axis is scaled by $\omega_{m+1} - \omega_m$. The curve labelled MFA is given by the mean field approximation described in the appendix.

## 5 Conclusions

In the spirit of Landauer's picture for electrons, we have developed a theoretical model for the thermal transport of ballistic phonons through a narrow bridge linking a thermal mass (the cavity) to the environment. We have obtained a simple relation between the thermal conductance of a single bridge mode and the mode cutoff frequency, which is analogous to the formula for quantized electronic conductance. Our simple picture indicates that, in contrast to the case for Fermi-distributed electrons, the thermal conductance of Bose distributed phonons does *not* exhibit dramatic features reflecting the underlying quantization of modes as the temperature is varied. This appears to hold even at the lowest temperatures when the thermal energy is smaller than the average mode cutoff spacing for both adiabatic and abrupt coupling between the narrow channel and the larger regions. This is a consequence of the fact that both the width and average energy of the Bose distribution are set by temperature, whereas for the Fermi distribution, the energy is determined by an independently controllable parameter, i.e. electron density.

We have studied non-equilibrium phonon distributions, where the effects of individual bridge modes become more evident. We find quantized features in the thermal decay when the phonon occupation occurs over a range narrower than a thermal distribution. Alternatively, in the situation where the cavity phonon distribution does not have sufficient time to rethermalize during the cooling an interesting phonon spectral density develops from an initial thermal distribution, directly reflecting the discrete mode structure of the thermal transport. The decay of the total energy in this case is exponential for ideal coupling between cavity and bridge. However an approximate model where transmission into the lowest mode of the channel is weak, which approximates the behavior due to geometric impedances, shows that the thermal decay curve exhibits a crossover to a slower exponential decay at low temperatures.

The effect of imperfect coupling between cavity and bridge modes can be understood in terms of a transmission coefficient that we calculate for an abrupt junction in a scalar model of the elastic waves. The changes from the perfect coupling (adiabatic) case are small except at very low temperatures where only the lowest few bridge modes are populated. However the asymptotic low temperature dependences are significantly changed due to the strong effect of the geometric mismatch on the long wavelength acoustic modes. In particularly the thermal cooling, which is linear in time for the adiabatic coupling, develops a $t^{-1/2}$ long time tail, and the low temperature thermal conductance changes from linear in temperature to $T^4$.

These results provide an initial, intuitive guide to the physics of mesoscopic phonon transport. With the recent advent of new experimental techniques to explore phonon transport in *insulating* mesoscopic systems, we look forward to new surprises from the measurements.

# 6 Acknowledgments

The authors would like to acknowledge John Worlock and Ron Lifshitz for illuminating discussions. MLR acknowledges partial support for this work from NSF grant DMR-9705411. The work of DEA was supported by the SURF 1997 program at Caltech.

# 7 Appendix

In this appendix we present the calculations of the abrupt junction in more detail, following the analysis of Szafer and Stone [10].

Assume a simple two dimensional geometry consisting of a rectangular cavity of transverse dimension $W$ connected to a rectangular bridge of transverse dimensions $w$. (In the general three dimensional case, if the cavity and bridge have the same thickness, there is no mixing of the $z$ modes, and the problem separates into a set of two dimensional problems, one for each $z$ mode.) Let $\chi_\alpha^c(y)$ and $\chi_m(y)$ be orthonormal transverse modes in the cavity and the bridge, respectively. (For clarity we will denote cavity mode indices by Greek letters, and bridge mode indices by Roman letters.)

The solutions to the wave equation take the form $\phi_m(x, y, t) = \chi_m(y)e^{i(kx-\omega t)}$ for the bridge, where $\omega^2 = \omega_m^2 + c^2k^2$ with $\omega_m = m\pi c/w$ is the cutoff frequency of the $m$th bridge mode, and has a similar form in the cavity with the cavity width $W$ replacing the bridge width $w$.

Consider a phonon incident on the interface from the cavity side, in the mode $\alpha$ of the cavity, and with longitudinal wave vector $k_\alpha^{(c)}$. The solutions in the cavity and bridge are :

$$\begin{array}{ll} \phi^{(c)} = \chi_\alpha^{(c)} e^{ik_\alpha^{(c)}x} + \sum_\beta r_{\alpha\beta}\chi_\beta^{(c)} e^{-ik_\beta^{(c)}x} & \text{cavity,} \\ \phi = \sum_m t_{\alpha m}\chi_m e^{ik_m x} & \text{bridge.} \end{array} \tag{14}$$

In the above equations, $k_m$ and $k_\beta^{(c)}$ are the wave vectors of the transmitted and reflected waves respectively. They are given by the energy conservation condition

$$\omega^2 = c^2 k_\alpha^{(c)2} + \omega_\alpha^{(c)2} = c^2 k_m^2 + \omega_m^2 = c^2 k_\beta^{(c)2} + \omega_\beta^{(c)2}. \tag{15}$$

Note that the sums over $m$ and $\beta$ include evanescent waves (imaginary $k$ or $k^{(c)}$) although only the propagating modes will contribute to the energy transport. The two solutions have to be matched at $x = 0$, which leads to the equations:

$$\begin{array}{l} \chi_\alpha^{(c)} + \sum_\beta r_{\alpha\beta}\chi_\beta^{(c)} = \sum_m t_{\alpha m}\chi_m, \\ k_\alpha^{(c)}\chi_\alpha^{(c)} - \sum_\beta r_{\alpha\beta}k_\beta^{(c)}\chi_\beta^{(c)} = \sum_m t_{\alpha m}k_m\chi_m. \end{array} \tag{16}$$

By integrating the first equation with $\chi_\beta^{(c)}$, and making use of the orthonormality relation $\int dy\chi_\alpha^{(c)}\chi_\beta^{(c)} = \delta_{\alpha\beta}$, we obtain

$$r_{\alpha\beta} = -\delta_{\alpha\beta} + \sum_m t_{\alpha m}a_{m\beta}, \tag{17}$$

where $a_{m\beta}$ is the overlap of cavity and bridge transverse functions $a_{m\beta} = \int dy \chi_\beta^{(c)} \chi_m$. Eq.(17) may be plugged in the second equation in Eq.(16), and the result is:

$$2k_\alpha^{(c)} \chi_\alpha^{(c)} - \sum_m \sum_\beta t_{\alpha m} a_{m\beta} k_\beta^{(c)} \chi_\beta^{(c)} = \sum_m t_{\alpha m} k_m \chi_m, \tag{18}$$

which, when integrated with $\chi_m$, yields

$$2k_\alpha^{(c)} a_{m\alpha} = \sum_n A_{nm} t_{\alpha n} + t_{\alpha m} k_m, \tag{19}$$

which is a system of equations that determine $t_{\alpha m}$. In Eq.(19) the kernel $A_{mn}$ is given by

$$A_{mn} = \sum_\beta a_{m\beta} a_{n\beta} k_\beta^{(c)}. \tag{20}$$

These equations may be solved numerically for the $t$'s. However there is a simple approximation [10] that provides an analytic form for the solution that is in extremely good agreement with the exact solution. The approximation derives from three important properties of $a_{m\alpha}$. Firstly, $a_{m\alpha} = 0$ unless $m$ and $\alpha$ have the same parity. In other words, even modes couple to even modes and odd modes to odd modes only. Secondly as a function of $\alpha$, $a_{m\alpha}$ is sharply peaked around $\alpha = mW/w$, the width of the peak being of order $W/w$. And thirdly, $a_{\alpha m}$ must satisfy the completeness relation

$$\sum_\alpha a_{m\alpha} a_{n\alpha} = \delta_{mn}. \tag{21}$$

The first two properties permit the key approximation, namely that $A_{mn} \propto \delta_{mn}$ (since the product of two functions peaked at different channels $m, n$ is very small and $A_{mn}$ is rigorously zero when $m$ and $n$ are different parity modes). Then we only need the diagonal part of $A$

$$A_{mm} = \sum_\beta a_{m\beta}^2 k_\beta^{(c)}, \tag{22}$$

which is simply a weighted average of the complex wave vector over the narrow range of reflected cavity modes for which $a_{m\beta}$ is significant. (Note $\sum_\beta a_{m\beta}^2 = 1$ by completeness).

In this case Eq.(19) separates into

$$2k_\alpha^{(c)} a_{m\alpha} = A_{mm} t_{\alpha m} + k_m t_{\alpha m} \tag{23}$$

and then

$$t_{\alpha m} = \frac{2k_\alpha^{(c)} a_{m\alpha}}{A_{mm} + k_m}. \tag{24}$$

The flux transmission probability from the wave vector $k_\alpha^{(c)}$ state of cavity mode $\alpha$ to bridge mode $m$ is given by

$$\mathcal{T}_{\alpha m} = |t_{\alpha m}|^2 \frac{k_m}{k_\alpha^{(c)}} = \frac{4k_\alpha^{(c)} k_m |a_{m\alpha}|^2}{(k_m + K_m)^2 + J_m^2}, \tag{25}$$

where $K_m = ReA_{mm}$ and $J_m = ImA_{mm}$. We observe that $t_{\alpha m} \sim a_{m\alpha}$, and from the form of the overlap $a_{m\alpha}$, to a good approximation only the cavity modes $\alpha$ satisfying $(m-1)W/w < \alpha < (m+1)W/w$ and having the same parity as $m$ contribute to the energy transport through the $m$th mode of the bridge.

Now populating the cavity modes $\alpha$ with the distribution $n(\omega)$, and summing over the energy flux due to the amplitude $t_{\alpha m}$ in each bridge mode $m$ noting that only cavity and bridge modes of the same frequency are coupled, leads to the simple modification of the heat transport

$$H_m^{(+)} = \frac{1}{2\pi} \int_{\omega_m}^{\infty} d\omega \, \hbar\omega \, n(\omega) \, \mathcal{T}_m(\omega), \tag{26}$$

where the "transport transmission coefficient" into the $m$th mode $\mathcal{T}_m(\omega)$ is defined (for $\omega > \omega_m$) by

$$\mathcal{T}_m(\omega) = \sum_{\alpha, \omega_\alpha^{(c)} < \omega} \frac{4k_\alpha^{(c)} k_m |a_{m\alpha}|^2}{(k_m + K_m)^2 + J_m^2} = \frac{4K_m k_m}{(k_m + K_m)^2 + J_m^2}, \tag{27}$$

with $k_m(\omega)$ and $k_\alpha^{(c)}(\omega)$ given by Eq.(15).

We must now use the explicit form of the transverse modes to evaluate $\mathcal{T}_m(\omega)$. A natural scalar model of the full elastic theory is to assume $\phi$ satisfies the wave equation. Since the full elastic theory will permit an "acoustic mode" with $\omega \to 0$ as $k \to 0$ an appropriate boundary condition would appear to be $\partial\phi/\partial y = 0$ at the $y$ boundaries (rather than $\phi = 0$ as in the electronic case). With this boundary condition we have explicitly for each polarization:

$$\chi_m(y) = \begin{cases} \sqrt{\frac{2}{w}} \cos(\frac{m\pi y}{w}) & m \text{ even} \\ \sqrt{\frac{2}{w}} \sin(\frac{m\pi y}{w}) & m \text{ odd} \end{cases} \tag{28}$$

$$\chi_\alpha^c(y) = \begin{cases} \sqrt{\frac{2}{W}} \cos(\frac{\alpha\pi y}{W}) & \alpha \text{ even} \\ \sqrt{\frac{2}{W}} \sin(\frac{\alpha\pi y}{W}) & \alpha \text{ odd} \end{cases} \tag{29}$$

with $m$ and $\alpha$ integers, with the special case $\chi_0(y) = 1/\sqrt{w}$ and $\chi_0^c(y) = 1/\sqrt{W}$. The $a_{m\alpha}$ are easily calculated, e.g. for $m$,$\alpha$ even:

$$a_{m\alpha} \simeq \sqrt{\frac{w}{W}} \left[ \frac{\sin\left(\frac{\alpha\pi w}{2W} - \frac{m\pi}{2}\right)}{\frac{\alpha\pi w}{2W} - \frac{m\pi}{2}} + \frac{\sin\left(\frac{\alpha\pi w}{2W} + \frac{m\pi}{2}\right)}{\frac{\alpha\pi w}{2W} + \frac{m\pi}{2}} \right] \simeq \sqrt{\frac{w}{W}} \frac{\sin\left(\frac{\alpha\pi w}{2W} - \frac{m\pi}{2}\right)}{\frac{\alpha\pi w}{2W} - \frac{m\pi}{2}} \tag{30}$$

(for $m, \alpha \neq 0$) where the second approximation is good for large $m$. The $a_{m\alpha}$ are indeed sharply peaked as a function of cavity mode number $\alpha$. The large $m$ approximation is essentially identical to the result in the electronic case. Since most of the weight is concentrated in the peak between $(m-2)\,W/w < \alpha < (m+2)\,W/w$ we evaluate $K_m + i\,J_m$ as the weighted average over this peak [11]

$$K_m + i\,J_m \simeq \frac{\sum_\alpha a_{m\alpha}^2 k_\alpha^{(c)}}{\sum_\alpha a_{m\alpha}^2}, \tag{31}$$

with both sums running over this range.

The function $\mathcal{T}_m\,(\omega)$ is plotted as a function of the dimensionless parameter $\delta$ defined by $\delta = (\omega - \omega_m)\,/\,(\omega_{m+1} - \omega_m)$ for different $m$ in figure 11. Curves rising from 0 to 1 within a fraction of the frequency change to the next bridge mode are found. This can be understood since $K_m + i\,J_m$ is the (weighted) average of the complex wave vectors of the narrow band of cavity modes that couple to the $m$th bridge mode (figure 10). As $\omega$ increases, $J_m$ becomes close to zero, and $K_m$ approaches $k_m$ so that $\mathcal{T}_m\,(\omega)$ approaches unity. Also plotted are the results based on a step function approximation to $a_{m\alpha}$ that Szafer and Stone call the "mean field approximation":

$$a_{m\alpha}{}^2 \simeq \frac{w}{W}\left(\theta(\alpha - (m-1)\frac{W}{w}) - \theta(\alpha - (m+1)\frac{W}{w})\right) \tag{32}$$

which satisfies the completeness relation for $a_{m\alpha}$. (There are exactly $2W/w$ modes of the cavity between bridge modes number $m-1$ and $m+1$. Out of these, only $W/w$ have the same parity as $a$, and thus couple to it, which justifies the factor $w/W$ in front of the expression.) The large $m$ curves approach this approximate form quite closely. The resulting heat transport $H_m^{(+)}$ due to the single bridge mode $m$ in this approximation is also plotted in figure 4. The resemblance between this curve and the one obtained for the adiabatic case, is striking. The exact abrupt coupling curve is slightly below the adiabatic curve due to the reflections at the junction which reduce the heat transport, but by a surprisingly small amount.

The difference from the adiabatic calculation becomes more important at low temperatures, where the lower modes of the bridge dominate. In the limit $T \to 0$, only the lowest mode of the bridge (the acoustic mode) contributes to the energy transport, so that we need only

$$H_0^{(+)} = \frac{1}{2\pi}\int_0^\infty d\omega\,\hbar\omega\,n(\omega)\mathcal{T}_0(\omega), \tag{33}$$

where $\mathcal{T}_0(\omega)$ may be obtained from Eq.(27). Because of the factor $n(\omega)$ only small $\omega$ will contribute to the integral for $T \to 0$ , so that we are interested in the behavior of $\mathcal{T}_0(\omega)$ for $\omega << c\pi/w$. We then have Eq.(22):

$$K_0 \simeq 2\frac{w}{W}\sum_{\beta=0}^{\omega\frac{W}{2\pi c}}(\frac{\omega^2}{c^2} - \beta^2\frac{4\pi^2}{W^2})^{1/2} = \frac{w}{4c^2}\omega^2 \tag{34}$$

replacing the sum by an integral for large $W$. In the same limit $J_0$ is some number which has no strong dependence on $\omega$. From Eq.(27), and taking into account the linear dispersion for the acoustic mode $k \sim \omega$, we arrive at the conclusion that in this limit $\mathcal{T}_0 \sim \omega^3$. This result, of course, depends on the detailed assumptions about the nature of the dispersion curves.

# References

[1] "*Spatial variation of currents and fields due to localized scatterers in metallic conduction*", R. Landauer, IBM J. Res. **1**, 223 (1957).

[2] "*Relation between conductivity and transmission matrix*", D.S. Fisher and P.A. Lee, Phys. Rev. **B23,** 6851 (1981).

[3] "*Quantized conductance of point contacts in a two-dimensional electron-gas*", B.J. van Wees, H. Van Houten, C.W.J. Beenakker, J.G. Williamson, L.P. Kouwenhoven, Phys. Rev. Lett. **60**, 848 (1988); "*One-dimensional transport and the quantization of the ballistic resistance*", D. Wharam, T.J. Thornton, R. Newbury, M. Pepper, H. Ahmed, J. Phys. C. **21**, L209 (1988).

[4] "*Hot Electrons and Energy Transport in Metals at mK Temperatures*", M.L. Roukes, M.R. Freeman, R.S. Germain, and R.C. Richardson, Phys. Rev. Lett. **55**, 422 (1985).

[5] "*The physics and fabrication of ultra-thin free-standing wires*", C. Smith, H. Ahmed, M.J. Kelly, and M.N. Wybourne, Superlattices and Microstructures **1**, 153 (1985).

[6] "*Electron heating effects in free-standing single-crystal GaAs fine wires*", A. Potts, M. J. Kelly, C. G. Smith, D. B. Hasko, J. R. A. Cleaver, H. Ahmed, D. C. Peacock, D. A. Ritchie, J. E. F. Frost, and G. A. C. Jones, J. Phys. **C2**, 1817 (1990); "*Thermal transport in free-standing semiconductor fine wires*", A. Potts, M. J. Kelly, D. B. Hasko,C. G. Smith, J. R. A. Cleaver, H. Ahmed, D. C. Peacock, J. E. F. Frost, D. A. Ritchie, and G. A. C. Jones, Superlattices and Microstructures **9**, 315 (1991).

[7] "*Acoustic waveguide modes observed in electrically heated metal wires*", by J. Seyler and M. N. Wybourne, Phys. Rev. Lett. **69**, 1427 (1992).

[8] "*Role of phonon dimensionality on electron-phonon scattering rates*", J.F. DiTusa, K. Lin, M. Park, M.S. Isaacson, and J.M. Parpia, Phys. Rev. Lett. **68**, 1156 (1992); "*An attempt to observe phonon dimensionality crossover effects in the inelastic scattering rate of thin free-standing aluminum films*", Y.K. Kwong, K. Lin, M.S. Isaacson, and J.M. Parpia, Jour. of Low Temp. Phys. **58**, 88 (1992).

[9] "*Direct thermal conductance measurements on suspended monocrystalline nanostructures*", T.S. Tighe, J.M. Worlock, and M.L. Roukes, Appl. Phys. Lett. **70**, 2687 (1997).

[10] "*Theory of Quantum Conductance through a Constriction*" A. Szafer and A.D. Stone, Phys. Rev. Lett. **62**, 300 (1989)

[11] There is actually a difficulty here that we have not resolved: although $|a_{m\alpha}|^2$ is quite strongly peaked over this range, the $\alpha^{-2}$ tails mean that the sum over $\alpha$ in the definition of $K_m + i J_m$ actually diverges weakly (logarithmically) for an infinite number of transverse modes $\alpha$. This does not happen in the calculation of Szafer and Stone, where the corresponding full expression eventually falls off as $\alpha^{-4}$ even though the approximate form has the identical $\alpha^{-2}$ behavior. Thus the zero derivative boundary conditions we have introduced to mimic the correct behavior of the acoustic mode, has also subtly, and unfortunately, changed the large $\alpha$ behavior. Except for modeling the acoustic mode, zero boundary conditions for $\phi$ would seem entirely adequate ($\phi$ then relates to the stress rather than the displacement), and the divergence disappears. It is not clear to use whether this difficulty is an artifact of the scalar model of the elastic waves, or a physical phenomenon that will survive in a full elastic theory treatment. The large $\alpha$ terms contribute only to the imaginary part $J_m$ which would go to a constant depending logarithmically on the cutoff for large $\omega$ rather than to zero as in our calculation using the truncated range of $a_{m\alpha}$.